# Properties of Baryonic, Electric and Strangeness Chemical Potential and Some of Their Consequences in Relatiistic Heavy Ion Collisions

Aram Z. Mekjian

Rutgers University, Department of Physics and Astronomy, Piscataway, NJ. 08854 & California Institute of Technology, Kellogg Radiation Lab 106-38, Pasadena, Ca 91125

**Abstract**
Analytic expressions are given for the baryonic, electric and strangeness chemical potentials which explicitly show the importance of various terms. Simple scaling relations connecting these chemical potentials are found. Applications to particle ratios and to fluctuations and related thermal properties such as the isothermal compressibility $\kappa_T$ are illustrated. A possible divergence of $\kappa_T$ is discussed.

## 1.Introduction

The behavior of hadronic matter at moderate to high temperature $T$ and density $\rho$ is being studied by heavy ion collisions. Moderate energy collisions, such as those done at National Superconducting Cyclotron at MSU and at GANIL, focus on the liquid/gas phase transition[1]. Much higher energy collisions[2] at CERN or BNL RHIC investigate the very hot and dense regions of phase space which probe a quark-gluon phase. This phase is then followed by an expansion to lower $\rho$ and $T$ where the colored quarks and anti-quarks form isolated colorless objects which are the well known particles whose properties are tabulated in ref[3]. Experiments at the proposed FAIR [4] will study the phase diagram of hadronic matter at a $T$ and $\rho$ which fills the gap between medium energy and high energy collisions. Statistical models [1,5-7] of such collisions assume an equilibrium is produced and its predictions are compared with experiment. Properties of the observed particles depend on chemical potentials associated with conservation laws such as baryon number $B$ and electric charge $Q$ conservation. At higher energies, when strange particles are also produced, strangeness conservation is also present in the strong interactions. This paper focuses on the behavior and properties of the three chemical potentials $\mu_B, \mu_Q, \mu_S$. Simple analytic expressions are developed for $\mu_B, \mu_Q, \mu_S$ which explicitly show the importance of various quantities that appear. These expressions are then used to study particle ratios, particle asymmetries such as the baryon/antibaryon asymmetry, fluctuations and thermal properties of hadronic matter. Results based on a Hagedorn resonance gas model are developed. A possible divergence of the isothermal compressibility is discussed. The critical exponent associated with this divergence is related to properties of the vanishing of chemical potentials with $T$ and the mass spectrum of excited states.

## 2. Properties of the Hadron Phase

*2.1 General Results*

The statistical model of relativistic heavy ion collisions [5-7] assumes that a frozen equilibrium is established in a volume $V$ and at a temperature $T$ and proceeds to explore

its consequences. The particle yields $< N_i >$ can be written in a simplified notation as

$$< N_i >= a_i x^{b_i} y^{q_i} z^{(-s_i)} \qquad (1)$$

in the non-degenerate limit. The $a_i = g_s(i)(VT^3/2\pi^2)(m_i/T)^2(K_2(m_i/T))$. Small mass differences between different charge states of the same particle will be ignored in $a_i$. The $g_s(i)$ is the spin degeneracy of particle $i$, which has mass $m_i$, baryon number $b_i$, charge $q_i$ and strangeness $s_i$. The $x \equiv \exp[\mu_B/T]$, $y \equiv \exp[\mu_Q/T]$ and $z \equiv \exp[-\mu_S/T]$ are determined by the constraints on baryon number $B$, charge $Q$ and strangeness $S$ that read: $B = = \Sigma b_I < N_i >$, $Q = \Sigma q_i < N_i >$, $S = \Sigma s_i < N_i >$. Because total $B, Q, S$ are each conserved so are the hypercharge $Y = B + S = \Sigma(b_i + s_i) < N_i >$ and the third component of isospin $I_Z$. The $I_Z$ is connected to the hypercharge $Y = B + S$ through the Gell-Mann Nishijima expression[8]: $2I_Z = 2Q - Y = \Sigma(2q_i - b_i - s_i) < N_i > = Z - N$. The $Z - N$ is determined by the initial proton/neutron asymmetry in the target and projectile. In a heavy ion collision the net strangeness is zero, but $\mu_s \neq 0$. The negative strangeness is from hyperons such as $\Lambda^0, \Sigma^+, \Sigma^0, \Sigma^-$ and anti-kaons $K^-, \overline{K}^0$, while the positive strangeness is basically in kaons $K^+, K^0$.

The simple quark model describes the hadrons considered here (strange and non-strange mesons and baryons) in terms of $u, d,$ and $s$ quarks and $\overline{u}, \overline{d},$ and $\overline{s}$ anti-quarks. Since baryons are restricted to three quarks and mesons to quark, anti-quark pairs, the quantum numbers of the hadrons are restricted to $b_i = \pm 1, 0$, $q_i = \pm 2, \pm 1, 0$, and $s_i = \pm 3, \pm 2, \pm 1, 0$. Contributions of excited states of a given particle can be added to the lowest members contribution so that $a_i \to A_i$. For instance, the number of $\Delta$ like particles $<\Delta(1232)> + <\Delta(1600)> + \ldots = A_\Delta xz(y^2 + y + 1 + 1/y)$ where $A_\Delta = a(\Delta(1232)) + a(\Delta(1600))\ldots$. The $A_\pi$ will contain $\pi, \rho, \ldots$ and $A_N$ has $p, n, N^*, \ldots$. Particles are grouped with $J = \{N, \Delta, \Lambda, \Sigma, \Xi, \Omega, \pi, K\}$ and $< N_J >= A_J x^{b_i} y^{q_i} z^{(-s_i)}$. An exponentially increasing Hagedorn density of states can be included as discussed below in sect. 2.3.

*2.2 Coupled equations for $x, y, z$ or the chemical potentials $\mu_B, \mu_Q, \mu_S$.*

The Gell-Mann/ Nishijima expression for $2I_Z = 2Q - Y$ as a constraint condition gives:

$$2I_Z = A_N x(y-1) + A_\Delta x(3y^2 + y - 1 - 3y^{-1}) + A_\Sigma xz(2y + 0y - 2y^{-1}) + A_\Xi xz^2(-y^{-1} + 1)$$
$$+ A_\pi(2y + 0y - 2y^{-1}) + A_K\{z(-y^{-1} + 1) - (1/z)(-y+1)\} - 2I_{W,Z}(\overline{B}) \qquad (2)$$

The $2I_Z(\overline{B})$, which is the anti-baryon contribution, is obtained by taking reciprocals of $x, y$ and $z$. Every numerical coefficient in front of $y$, when divided by 2, is just the isospin $I_Z$ component of each charged state of particle type $J$.

For $S = 0$, $2I_Z = Z - N$ and for symmetric $N = Z$ systems $I_Z = 0$. Eq.(2) explicitly shows that when $I_Z=0$ then $y=1$. At $y=1$, the $\mu_Q=0$. For an asymmetric system $N \neq Z$ and thus $y \neq 1$. Usually $/\mu_Q/<<T$ and $\mu_Q<0$ since typically $N>Z$ and $(N-Z)/B<<1$. A lowest order expansion around $y=1$ gives $\mu_Q \sim (Z-N)$ with

$$\frac{\mu_Q}{T} = \frac{2I_Z = 2Q-Y = Z-N}{(A_N + 10A_\Lambda)(x+\frac{1}{x}) + 4A_\Sigma(xz+\frac{1}{xz}) + A_\Xi(xz^2+\frac{1}{xz^2}) + 4A_\pi + A_K(z+\frac{1}{z})} \quad (3)$$

from eq.(2). Three terms in the denominator of eq.(3) are important. The large factor of 10 in $10A_\Lambda$ makes $\Lambda$ central in $\mu_Q/T$. The abundance of pions coming from the collision results in a large contribution through the $4A_\pi$ factor. The initial protons in the target and projectile are contained in the $A_N x$ factor. Of much less importance are the strange particle contributions in $\mu_Q$. The evaluation of $x,z$ for $I_Z=0$ where $y=1$ is given next. As a first approximation these values also apply when $I_Z \neq 0$ since $/\mu_Q/$ is usually $<<\mu_S,\mu_B$. The $B = N_B - N_{\bar{B}}$ constraint equation reads:

$$N_B = x\{A_N(y+1) + A_\Lambda(y^2+y+1+\frac{1}{y}) + A_\Lambda z + A_\Sigma z(y+1+\frac{1}{y}) + A_\Xi z^2(\frac{1}{y}+1) + A_\Omega \frac{z^3}{y}\}$$
$$\rightarrow (y=1) \rightarrow x\{2A_N + 4A_\Lambda\} + A_\Lambda z + 3A_\Sigma z + 2A_\Xi z^2 + A_\Omega z^3\} \equiv xC(z) \quad (4)$$

The coefficient $\{(2A_N + 4A_\Lambda) + A_\Lambda z + 3A_\Sigma z + 2A_\Xi z^2 + A_\Omega z^3\} \equiv C(z)$ is a polynomial in $z$. The $N_{\bar{B}}$ is obtained from $N_B$ by taking the reciprocal of $x,y$ and $z$ so that $N_{\bar{B}} = C(1/z)/x$. Given that $x$ only appears in $N_B$ and $1/x$ in $N_{\bar{B}}$, a quadratic equation in $x$ arises from the $B$ constraint equation which reads $B = C(z)x - C(1/z)/x$ or $C(z)x^2 - Bx - C(1/z) = 0$. Thus $x = (B + \sqrt{B^2 + 4C(z)C(1/z)})/2C(z)$. Since each $A_J \sim V$, $x = f(B/V,T)$. The result of eq.(4) can be substituted into: $-S = N_{S-} - N_{S+} = 0$ to find $z$. The $N_{S-}$ is:

$$N_{S-} = x\{A_\Lambda z + A_\Sigma z(y+1+1/y) + 2A_\Xi z^2(1/y+1) + 3A_\Omega z^3/y\} + A_K z(1/y+1)$$
$$\rightarrow (y=1) \rightarrow x\{A_\Lambda z + 3A_\Sigma z + 4A_\Xi z^2 + 3A_\Omega z^3\} + 2A_K z \quad (5)$$

The $N_{S+}$ is obtained by taking the reciprocal of $x, y$ and $z$. The solutions to the resulting polynomial equation for $z$ simplify in certain limits which are now discussed.

2.2 Simplified solutions in symmetric $N=Z$ systems at $y=1$ and $\mu_S/T <<1$ or $z \approx 1$.

When $\mu_S/T << 1$, $z = 1 - \mu_S/T$, and $\mu_S/T$ is given by

$$\mu_S/T = \sinh(\mu_B/T)H_{S,1}/1^2(2A_K) + \cosh(\mu_B/T)H_{S,2} \quad (6)$$

The $H_{S,j} = 1^j(1A_\Lambda) + 1^j(3A_\Sigma) + 2^j(2A_\Xi) + 3^j(1A_\Omega)$ are strangeness moments of the hyperon

distribution. If $\mu_B/T \ll 1$, $\cosh(\mu_B/T) \to 1$, $\sinh(\mu_B/T) \to \mu_B/T$, and

$$\mu_S/T = (\mu_B/T)H_{S,1}/(1^2(2A_K) + H_{S,2}) \equiv C_{SB}\mu_B/T \tag{7}$$

The slope of $\mu_S$ versus $\mu_B$ involves the magnitude of the strangeness in hyperons to the strangeness fluctuation including kaons. When $x \approx 1, z \approx 1, y \approx 1$, the $< N_i > \approx a_i$. A factor similar to $C_{SB}$ appears in the discussion of a $B/S$ correlation in ref.[9]. For large $\mu_B/T$, $\{\sinh(\mu_B/T), \cosh(\mu_B/T)\} \to \exp(\mu_B/T)/2$. Substituted this result into Eq.(6) gives a non-linear relation between $\mu_S/T$ and $\mu_B/T$. The $\mu_B/T$ satisfies:

$$\frac{B}{2} = \sinh(\frac{\mu_B}{T})(2A_N + 4A_\Delta + \frac{2A_K H_{S,0} + (H_{S,0}H_{S,2} - (H_{S,1})^2)\cosh(\mu_B/T)}{2A_K + H_{S,2}\cosh(\mu_B/T)}) \tag{8}$$

At low $T$, where $x \gg 1$, and $z$ is small, the main contribution comes from $S = \pm 1$ strange particles ($K, \Lambda, \Sigma$). Without $\Xi, \Omega$ terms, the relation between $\mu_S/T, \mu_B/T$ is now

$$\tanh(\mu_S/T) = \sinh(\mu_B/T)\frac{1(A_\Lambda + (3A_\Sigma))}{1^2(2A_K) + \cosh(\mu_B/T)[1^2(A_\Lambda + (3A_\Sigma))]} \tag{9}$$

With $\mu_B/T \gg 1$, $\cosh(\mu_B/T)$ and $\sinh(\mu_B/T) \to \exp(\mu_B/T)/2$. The hypercharge equation $Y = (2A_N + 4A_\Delta)2\sinh(\mu_B/T) + 2A_K 2\sinh(\mu_S/T)$ gives $\mu_B/T$ in closed form.

*2.3 Role of the Hagedorn mass spectrum, vanishing of chemical potentials and $B, \bar{B}$ asymmetries.*

The Hagedorn density of excited states is $\rho = D_\tau m^{-\tau}\exp(\beta_h m)$, where $D_\tau$ is a constant and $\beta_h = 1/T_0$. The $T_0$ is the limiting $T$ and the exponent $\tau$ is a parameter. This density will affect $A_J$ which are sums over the lowest plus all exited states of $J$. For a Hagedorn density the $A_J \sim V\ (y/m_{J0})^{\tau-5/2}(\int_y^\infty(dxe^{-x}/x^{(\tau-3/2)}))$ with $y = (T_0 - T)m_{J0}/T_0 T$. The $m_{J0}$ is the lowest mass of particles of type $J$. For $\tau < 5/2$, $A_J \to \infty$ as $1/(T_0 - T)^{(5/2-\tau)}$. An infinite $A_J$ will result in $\mu_B, \mu_S, \mu_Q \to 0$ and the associated fugacities $x, y, z \to 1$. As a simplified example, a system with one chemical potential $\mu_B$ has:

$$(\frac{N_B - N_{\bar{B}}}{V})(\frac{2\pi}{T})^{3/2} = 2D_\tau(\frac{y}{m_{J0}})^{\tau-5/2}\left[\int_y^\infty(dxe\frac{e^{-x}}{x^{\tau-3/2}})\right]\sinh(\mu_B/T) \tag{10}$$

For $\tau < 5/2$, $\mu_B$ will go to zero as $(T_0 - T)^{(5/2-\tau)}$, when $T \to T_0$, the Hagedorn limiting $T$. The behavior of $I_{y,\tau} \equiv \int_y^\infty(dxe^{-x}/x^{(\tau-3/2)})) \to \Gamma(5/2 - \tau)$ with $\tau$ for $y \to 0$ is essential for determining the properties of $A_J$ and $\mu_B$. For $\tau > 5/2$, $\mu_B$ will go to a constant that depends on $\tau$. For $\tau = 5/2 + \eta$, $(y)^{\tau-5/2}\int_y^\infty dxe^{-x}/x^{(\tau-3/2)} = 1/\eta$ as $y \to 0$. When $5/2 < \tau < 7/2$, the constant $1/\eta$ is approached with $\infty$ slope: $y^{1/2} I_{y,3} = 2(1-(\pi y)^{1/2})$. For $\tau > 7/2$, the slope is finite. The $(N_B + N_{\bar{B}})/V \sim y^{\tau-5/2} I_{y,\tau} \cosh(\mu_B/T)$ which diverges as $1/(T_0-T)^{5/2-\tau}$ for $\tau < 5/2$ and for $\tau \to 5/2^-$ this quantity $\to -\ln(m_0(T_0-T)/T_0^2)$. The asymmetry $Asy(B\bar{B}) = (N_B + N_{\bar{B}})/(N_B - N_{\bar{B}}) = \coth(\mu_B/T) = (x+1/x)/(x-1/x)$, with $N_B - N_{\bar{B}} = B$ fixed. An $\infty$ density presents a problem if particles are not point like. The model is thus limited to $\rho$ and $T$ where composite baryons and mesons don't overlap. At some value $\rho \sim 1 hadron/fm^3$ and $T = T_{Qg} < T_0$, a transition to a quark-gluon phase occurs which truncates behaviors based on this particular model. A discussion of this feature can be found in ref[10] where the Qg phase is treated in a statistical model of Qg bags.

In [6], data are fit to a statistical model with $\mu_B = 80.85(T_0-T)^{1/2}$ in $MeV$ and $T_0 = 167$. A $(T_0-T)^{1/2}$ dependence occurs for $\tau = 2$. This form for $\mu_B$ will be used. Then:

$$x = \exp(\mu_B/T) = \exp[(80.85/T)(T_0-T)^{1/2}] \tag{11}$$

Because of the large cost factor $80.85 T_0^{1/2}/T = 1045/T \sim m_N/T$, large $Asy(B\bar{B})$ occur only if $T \approx T_0$. At $T = .99T_0$, $Asy(B\bar{B}) \approx 2$ and at $T = .9T_0$, $Asy(B\bar{B}) = 1.06$. When $T \gg \mu_B$, $Asy(B\bar{B}) \to T/\mu_B$.

Given $x$ or $\mu_B$, the $\mu_S$ and $z$ follow from results noted above while eq.(3) gives $\mu_Q$ and $y$. Simple scaling laws relating $\mu_S$ and $\mu_Q$ to $\mu_B$ are developed in sect.2.5.

*2.4 Particle ratios and fluctuations*

Results from the previous section can be used to study particle ratios prior to resonance decays. The final distributions of particles seen experimentally are changed by decays. A few examples are now presented which will illustrate the importance of various factors. When $y = 1$, $K^+/K^- = \exp(2\mu_S/T)$. The $K^+/K^-$ ratio can also be used to discuss the role of dropping masses [11] which leads to an enhancement of this ratio. The simplest result for this ratio is at low $T$ when $\mu_B/T \gg 1$ and when $\Xi, \Omega$ are neglected. Then

$$\frac{K^+}{K^-} = \exp(2\mu_S/T) = \frac{1}{z^2} \approx 1 + \exp(\frac{\mu_B}{T}) \frac{2m_\Lambda^2 K_2(m_\Lambda/T) + 6m_\Sigma^2 K_2(m_\Sigma/T)}{2m_K^2 K_2(m_K/T)}$$

$$\approx 1 + \exp(\frac{\mu_B - m_\Lambda + m_K}{T}) \frac{m_\Lambda^{3/2}}{m_K^{3/2}} \{1 + 3\frac{m_\Sigma^{3/2}}{m_\Lambda^{3/2}} \exp(\frac{-(m_\Sigma - m_\Lambda)}{T})\} \tag{12}$$

The exponential factor $\mu_B - m_\Lambda + m_K$ plays a significant role in $K^+/K^-$. For $N > Z$ $y = \exp(\mu_Q/T) < 1$ and the $K^+/K^- = \exp(2\mu_S/T)\exp(2\mu_Q/T)$ is reduced. When $\mu_S$ and $\mu_B$ are connected by eq.(7), the kaon asymmetry $(K^+ - K^-)/K^- = 2\mu_S/T = 2C_{SB}\mu_B/T$. The $C_{SB}$ is the baryon strangeness correlation coefficient involving the strangeness in hyperons to the strangeness fluctuation in kaons plus hyperons-see eq.(3).

The $\pi^-/\pi^+$ ratio $=\exp(-2\mu_Q/T)$. A good approximation (~1% error at T=120Mev) is:

$$\frac{\pi^-}{\pi^+} \approx 1 + 2\left(\frac{N-Z}{B}\right)\frac{(2A_N + 4A_\Delta)(x+1/x)}{(A_N + 10A_\Delta)(x+1/x) + 4A_\pi} \quad (13)$$

The $(\pi^- - \pi^+)/\pi^+$ charge asymmetry ratio prior to resonance decays $\sim (N-Z)/B$. This ratio also involves the $T$ dependent amplitudes: $A_N, A_\Delta, A_\pi$, with the numerator containing only the baryon amplitude $A_N, A_\Delta$ while the denominator also includes the meson amplitude $A_\pi$. This remark is similar to that for kaons which involved $C_{SB}$. Anti-baryons appear as $1/x$ terms and are small until $T \approx T_0$. At a low enough $T$ where $A_N x$ dominates, the $\pi^-/\pi^+ \to 1 + 4(N-Z)/B$ which is a limiting value for $\pi^-/\pi^+$.

The fluctuation in $\Phi$ is $\delta\Phi^2 = <\Phi^2> - <\Phi>^2$. In the grand canonical ensemble $\delta\Phi^2 = T(\partial<\Phi>/\partial\mu_\phi) = \Sigma\phi_i^2 <N_i> = \Sigma\phi_i^2 a_i x^{b_J} y^{q_J} z^{-s_J}$ where $\phi_i$ is the quantum number associated with particle $i$. The fluctuations in strangeness, where $\Phi = S$ and $\mu_\phi = \mu_S$, were already shown to be important in the relation between $\mu_S$ and $\mu_B$ as given in eqs.(6,7). The fluctuations in baryon number (where $\Phi = B$ and $\mu_\phi = \mu_B$) is simply $N_B + N_{\bar{B}}$ since $b_i = \pm 1$ only. In a Hagedorn model this fluctuation can diverge. Such divergences are related to singularities in the baryonic compressibility $\kappa_{T,B} = -(1/V)(\partial V/\partial P_B)_T$. The $(\partial P_B/\partial V)_T = (N_B - N_{\bar{B}})(\partial\mu_B/\partial V)_T/T$ from $P_B V = (N_B + N_{\bar{B}})T$ and $(V/T)(\partial\mu_B/\partial V)_T = -(N_B - N_{\bar{B}})/(N_B + N_{\bar{B}})$ from $B$ conservation giving:

$$\frac{<B>}{V}T\kappa_{T,B} = \frac{(<B^2> - <B>^2)}{<B>} = \frac{N_B + N_{\bar{B}}}{N_B - N_{\bar{B}}} = Asy(B\bar{B}) = -1/(\frac{V}{T}(\frac{\partial\mu_B}{\partial V})_T) \quad (14)$$

From sec.(2.3), $(N_B + N_{\bar{B}})$ diverges as $1/(T_0 - T)^{5/2-\tau}$ for $\tau < 5/2$, while $B$ is fixed at $<B> = (N_B - N_{\bar{B}})$. A critical exponent $\gamma$ is associated with the divergence of $\kappa_T$, with $\kappa_T \sim 1/(T_0 - T)^\gamma$. Consequently, the critical exponent $\gamma$ and the prefactor exponent $\tau$ are connected by $\gamma = 5/2 - \tau$. In turn, $\gamma$ is connected to the exponent in the vanishing behavior of $\mu_B \sim (T_0 - T)^{5/2-\tau}$. The compressibility of a pure gas of massless pions with $\mu = 0$, including degeneracy terms, is [12] $\kappa_{T,\pi} = (\varsigma(2)/\varsigma(3))V/(<N_\pi>T)$ which is somewhat larger than an ideal gas result $\kappa_T = 1/P = V/(<N>T)$ because of the statistical attraction of bosons which increases the compressibility. The zeta functions that appear in this expression for $\kappa_T$ arise from degeneracy corrections. The ideal gas $\kappa_T$ follows from eq.(14) when $\delta B^2 = <B>$, which is the poissonian limit. Then the rhs in the first equality in eq.(14) is 1 and $\kappa_{T,B} = V/(<B>T)$. By constrast, the $\kappa_T$ of a van der Waals gas becomes infinite at a critical point in a liquid gas phase transition which is responsible for the phenomena of critical opalescence. In such a phase transition the density fluctuations become large since the system doesn't know whether to be in a gas phase or a liquid phase.

Properties of $\mu_B$ are also reflected in the baryonic heat capacity $C_{V,B}$ at constant $V$ [13]:

$$\frac{C_{V,B}}{N_B + N_{\bar{B}}} = -[\coth(\frac{\mu_B}{T})] \cdot T \frac{\partial^2 \mu_B}{(\partial T)^2} + [\csch^2(\frac{\mu_B}{T})] \cdot T(\frac{\partial \mu_B}{\partial T} - \frac{\mu_B}{T})^2 \quad (15)$$

Here, the derivatives of $\mu_B$ are with respect to $T$ at constant $V$. The singularities of $C_{V,B}$ are discussed in ref.[13]. The importance of the difference of charge fluctuations in the hadron phase and in the quark gluon phase were discussed in ref.[14,15].

*2.5 Scaling laws*

Using eq.(11) for $\mu_B$, values of $\mu_S$ and $\mu_Q$ follow from the constraint equations. When $\Xi, \Omega$ contributions are neglected, then eq.(12) is a simple connection of $\mu_S$ to $\mu_B$ which can be rewritten as $\tanh(\mu_S/T) = f/(1+f)$ with $f = (1/2)\exp(\mu_B/T)R(T)$. The $R(T)$ is the ratio of mass terms that appear in eq.(12) with $=(a_\Lambda + 3a_\Sigma)/2a_K$. For $T < 100 MeV$ a scaling relation $\mu_S/\mu_B \approx .19 \approx 1/5$ is found. The behavior of $\mu_S/\mu_B \approx .2 - (.02T/100)$ is a good approximation for $0 < T < 160$ without $\Xi, \Omega$ contributions. The $\Xi, \Omega$ terms become significant at $T \sim 120 MeV$. The scaling relation $\mu_S/\mu_B \approx 1/5$ is still a good approximation with $\Xi, \Omega$ terms. Very near $T_0$, eq.(7) gives $\mu_S/\mu_B \approx .20 = 1/5$ and contains not only $\Xi, \Omega$, but also anti-hyperons. Including excited states of hyperons enhances $\mu_S$ while $K^*$'s decrease $\mu_S$ since $(a_\Lambda + 3a_\Sigma)/a_K \to (A_\Lambda + 3A_\Sigma)/2A_K$ in $f$.

The $\mu_Q$ follows from eq.(3) and the behavior of $\mu_Q$ is dominated by the $N, \Delta, \pi$ contributions. A scaling law $\mu_Q = -((N-Z)/B)\mu_B/10$ also applies. Thus:

$$\mu_B \approx 80.85(T_0 - T)^{1/2} MeV, \quad \mu_S \approx (15.36 - 16.17) MeV(T_0 - T)^{1/2},$$
$$\mu_Q \approx -((N-Z)/B)\mu_B/10 = -((N-Z)/B)\, 8.085\, MeV(T_0 - T)^{1/2} \quad (17)$$

The coefficient 15.36 is when $\mu_S/\mu_B \approx .19$ while 16.17 is with 1/5=0.2. Once $\mu_B, \mu_S$ and $\mu_Q$ determined, the $< N_i/V >$ then follow from eq.(1).

**3. Summary and Conclusions**

Properties of the three chemical potentials $\mu_B, \mu_Q, \mu_S$ were studied using an approach that lead to analytic expressions for them. The Gell-Mann/Nishijima expression for the third component of isospin $2I_Z = 2Q - Y$, where the hypercharge $Y = B + S$, was used to obtain the electric chemical potential $\mu_Q$. The $\mu_Q$ was then shown to depend on the initial proton/neutron asymmetry $(Z - N)/(Z + N)$. The abundance of initial nucleons, the large production of charged pions and $\Delta$'s with several charged states plays an essential role in determining the value of $\mu_Q$, besides the asymmetry factor $(N-Z)/(N+B)$. The $\mu_Q$ was also shown to scale with $\mu_B$ as $\mu_Q \approx -((N-Z)/B)(\mu_B/10)$. Simple connections between the strangeness $\mu_S$ and baryon $\mu_B$ were developed. A main element in this connection was moments of the distribution of strangeness. Specifically, the ratio of strangeness carried in hyperons to the strangeness fluctuation in hyperons plus kaons determined the slope of $\mu_S$ with $\mu_B$ at high $T$. For $T \approx T_0$ $\mu_S / \mu_B \approx 1/5$ where the 1/5 reflects this strangeness ratio. At low $T$, where $\Xi, \Omega$ contribution can be neglected, a non-linear equation between $\mu_S$ and $\mu_B$ was also obtained. Moreover, solutions to this non-linear equation also show a scaling relation is present that reads $\mu_S / \mu_B \approx 1/6$ valid for $T < 140 MeV$.

The role of the Hagedorn mass spectrum in the temperature dependence of chemical potentials was investigated. The prefactor exponent, labeled $\tau$, plays an essential role in the temperature dependence of chemical potentials determining whether they vanish as the critical Hagedorn $T_0$ is approached. A large baryon/anti-baryon asymmetry, as measured by $(N_B + N_{\bar{B}})/(N_B - N_{\bar{B}})$, exists only if $\tau \leq 5/2$ and $T \approx 0.98 T_0$. The result that $T$ has to be within a few % of $T_0$ has its origin in the high cost associated with $m_N / T$. The expressions developed for $\mu_B, \mu_Q, \mu_S$ were used to study particle ratios, such as the pion charge asymmetry $(\pi^- - \pi^+)/\pi^+$, the kaon strangeness ratio $K^+/K^-$, prior to adjustments from resonant decays. Terms important in these ratios and asymmetries were studied. For example, $\mu_B - m_\Lambda + m_K$ is important in $K^+/K^-$. Fluctuations were also investigated and connected to associated thermal quantities such as the relation of the isothermal compressibility to particle number fluctuations. A relation $5/2 - \tau = \gamma$ between the prefactor exponent $\tau$ and the critical exponent $\gamma$, which determines the divergence of the isothermal compressibility, was noted.

This work was supported in part by the DOE grant number DE-FG02-96ER-40987.